\documentclass[10pt]{iopart}
\usepackage{cite}

\expandafter\let\csname equation*\endcsname\relax
\expandafter\let\csname endequation*\endcsname\relax
\usepackage{amsmath}
\usepackage{color}
\usepackage{graphicx}
\usepackage{bm}
\usepackage{hyperref}

\begin{document}
\title{\emph{Ab initio} theory of the impact of grain boundaries and substitutional defects on superconducting Nb$_3$Sn}
\author{Michelle M Kelley, Nathan S Sitaraman and Tom\'as A Arias}
\address{%
 Department of Physics, Cornell University, Ithaca, New York 14853, USA
}%
\ead{mmk255@cornell.edu}
\date{\today}
\begin{abstract}
 Grain boundaries play a critical role in superconducting applications of Nb$_3$Sn: in dc applications, grain boundaries preserve the material's inherently high critical current density by pinning flux, while in ac applications grain boundaries can provide weak points for flux entry leading to significant dissipation. We present the first \emph{ab initio} study to investigate the physics of different grain boundary types in Nb$_3$Sn and their impact on superconductivity using density-functional theory. We identify an energetically favorable selection of high-angle tilt and twist grain boundaries of distinct orientations. We find that clean grain boundaries free of point defects reduce the Fermi-level density of states by a factor of two, an effect that decays back to the bulk electronic structure $\sim$1--1.5~nm from the boundary. We further calculate the binding free-energies of tin substitutional defects to multiple boundaries, finding a strong electronic interaction that extends to a distance comparable to that of the reduction of density of states. Associated with this interaction, we discover a universal trend in defect electronic entropies near a boundary. We then probe the effects of defect segregation on grain boundary electronic structure and calculate the impact of substitutional impurities on the Fermi-level density of states in the vicinity of a grain boundary. We find that titanium and tantalum defects have little impact regardless of placement, whereas tin, copper, and niobium defects each have a significant impact but only on sites away from the boundary core. Finally, we introduce a model for a local superconducting transition temperature and consider how grain boundary composition affects $T_\textrm{c}$ as a function of distance from the boundary plane. The methodology established in this manuscript can be applied to other A15 superconductors in addition to Nb$_3$Sn.
\end{abstract}
\noindent{\it Keywords\/}: Nb$_3$Sn, grain boundaries, density functional theory,  electronic structure, superconductivity, A15 superconductors, local $T_\textrm{c}$
\maketitle

\ioptwocol
\section{\label{sec:intro}Introduction}
Conventional niobium-based superconductors continue to dominate industrial superconducting technologies, and with alluring prospects of high-$T_\textrm c$ superconductors stunted by their disappointing transport properties, further optimization of Nb$_3$Sn in particular may be the most promising path towards higher performing devices \cite{Jin1991, Godeke2006,Foltyn2007, Lee2008, Xu2017}. While niobium-titanium alloys are the most widely-used superconductor, especially for high-field magnets in particle accelerators, Nb$_3$Sn offers the potential to operate both at higher fields and temperatures \cite{Garber1986, Ballarino2015}. However, many applications have historically avoided Nb$_3$Sn, despite its superior superconducting properties, considering the challenges involved in manufacturing the material in a way that preserves its desirable properties. The vitality of industrial applications of Nb$_3$Sn was made possible through the pioneering research of Dr. Suenaga and his colleagues at Brookhaven National Lab starting in the mid-1970s, notably through their research examining microstructural details that impact the material's superconducting performance and in particular, the studies revealing the importance of grain boundaries \cite{Farrell1974, Farrell1975, Suenaga1981,Suenaga1983,Tafto1984,Suenaga1985, Suenaga1985ii, Suenaga1986}. Since then, Nb$_3$Sn has been used or is actively being developed for superconducting applications in particle accelerators, fusion reactors, nuclear magnetic resonance instruments, and for non-magnet applications such as superconducting radio frequency (SRF) cavities \cite{Parrell2005,Ciazynski2007, sharma2015superconductivity, Posen2015, Ambrosio2015}. While Nb--Ti is recognized to be performing close to its intrinsic limits, superconducting applications adopting Nb$_3$Sn continue to improve, but the material's success hinges on controlling defects that impact performance and, moreover, a microscopic theory to predict the impact of competing defects known to be fundamental to the material's performance since the mid-1970s.

Nb$_3$Sn is a type-II superconductor belonging to the A15 class of superconductors which held the record for highest $T_\textrm{c}$ from 1954--1986~\cite{Echarri1971, Dew-Hughes1975, Flukiger1987, G.R.Stewart2015}. With a critical temperature of 18~K, Nb$_3$Sn lives in an interesting regime sharing features of both elemental and high-$T_\textrm{c}$ superconductors. Grain boundaries are known to degrade the superconducting properties of cuprates, iron-based superconductors, and other modern high-$T_\textrm{c}$ superconductors, providing structural disorder on length scales comparable to their notably short coherence lengths~\cite{Babcock1995, Dimos1990, Hilgenkamp2002, Gurevich2011, Katase2011,Durrell2011,Iida2020}. Nb$_3$Sn is a conventional superconductor based on the elemental superconductor niobium, but unlike niobium, which has a coherence length of $\sim$50~nm, Nb$_3$Sn has a coherence length of $\sim$3~nm, approaching the scale of structural disorder from grain boundaries~\cite{Godeke2006, Draskovic2013, Posen2017}. There is even speculation that grain boundaries in Nb$_3$Sn may exhibit Josephson-like effects~\cite{Gurevich1992, Hall,Sheikhzada2014}.

Optimization of Nb$_3$Sn for superconducting applications hinges on achieving ideal grain boundary behavior. Grain boundaries in Nb$_3$Sn are fundamental for growing the material itself \cite{Farrell1974, Farrell1975, Togano1979, Besson2007, Santra2015}, and prevailing limitations spotlight differing roles of grain boundaries in dc vs. ac applications. In dc applications such as superconducting wires, high grain boundary densities provide needed pinning centers to prevent vortices from limiting the critical current density~\cite{Scanlan1975, Xu2017, Sandim2013}. In ac applications such as SRF cavities, significant dissipation arises once flux penetrates the cavity surface, and grain boundaries can lower the barrier for flux penetration~\cite{Posen2017, Transtrum2011, Liarte2017, Carlson2020}. The target field of use also needs to be considered when optimizing the material, governing whether to prioritize grain boundary size or composition as both are observed to affect transport properties~\cite{Lee2001, Lee2005, Lee2008}.

Disorder and internal variations in stoichiometry in Nb$_3$Sn are known to have a profound effect on the material's properties~\cite{Godeke2006, Kate2017, Flukiger2008}, and stoichiometric variations at grain boundaries has been observed experimentally~\cite{Suenaga1983, Sandim2013, Lee2020}. Numerical simulations based on empirical inputs find that strong gradients of tin composition can reduce the critical current density, the flux pinning force density, and flux pinning scaling field in superconducting wires \cite{Cooley2004, Li2017, Baumgartner2018}. These simulations have been successful at describing experimental data and mapping back onto empirical laws, but there still lacks a microscopic theory on the role of grain boundaries in this material. Developing a microscopic description that readily contrasts effects caused by clean grain boundaries from effects caused by grain boundaries with stoichiometry gradients could advance applications of this material by identifying specific defect characteristics to control more precisely.

Nb$_3$Sn is often engineered with ternary elements for various optimizations. Copper helps Nb$_3$Sn growth by lowering the A15 formation temperature, and titanium and tantalum raise $H_{\textrm c2}$ and prevent the transformation into the tetragonal phase~\cite{Godeke2006,GOLDACKER1985, Suenaga1986, Flukiger2008, Flukiger2008_2}. These impurities have been observed at and around grain boundaries with many reporting pronounced segregations of copper and titanium at grain boundaries~\cite{Suenaga1983, Rodrigues1995, Suenaga2007, Sandim2013}. Experimentally, we can see only how atomic composition varies locally and cannot directly see local deviations in the electronic properties. Due to experimental limitations, the impact of different defects on the local superconducting properties near grain boundaries is an area hardly explored \cite{Marken1986}. As Nb$_3$Sn continues to be optimized, distinguishing the effects among defects opens up exciting prospects for grain boundary engineering, for advances in artificial pinning centers, and for developments in promising precipitates to add to the material~\cite{Dietderich1997,Dietderich1998, Rodrigues2007, Raabe2014, Xu2014, Baumgartner2015, Xu2019}. No comprehensive study of grain boundaries in Nb$_3$Sn using density-functional theory currently exists, and \emph{ab initio} calculations would allow us to extend our understanding beyond the limits of analyses which rely upon general empirical laws.

This work presents a study of the influence of grain boundaries on the properties of Nb$_3$Sn from first principles using density-functional theory. We identify a selection of energetically favorable structures featuring both tilt and twist grain boundaries, study the impact of boundaries on the electronic structure, and investigate the behavior of point defects near a boundary. Motivated by recent experiments that found highly degraded quality factors in Nb$_3$Sn SRF cavities with excess tin at grain boundaries, we contrast how the local superconducting properties are impacted near a clean grain boundary versus a grain boundary containing their measured tin concentration profile~\cite{Lee2020}.

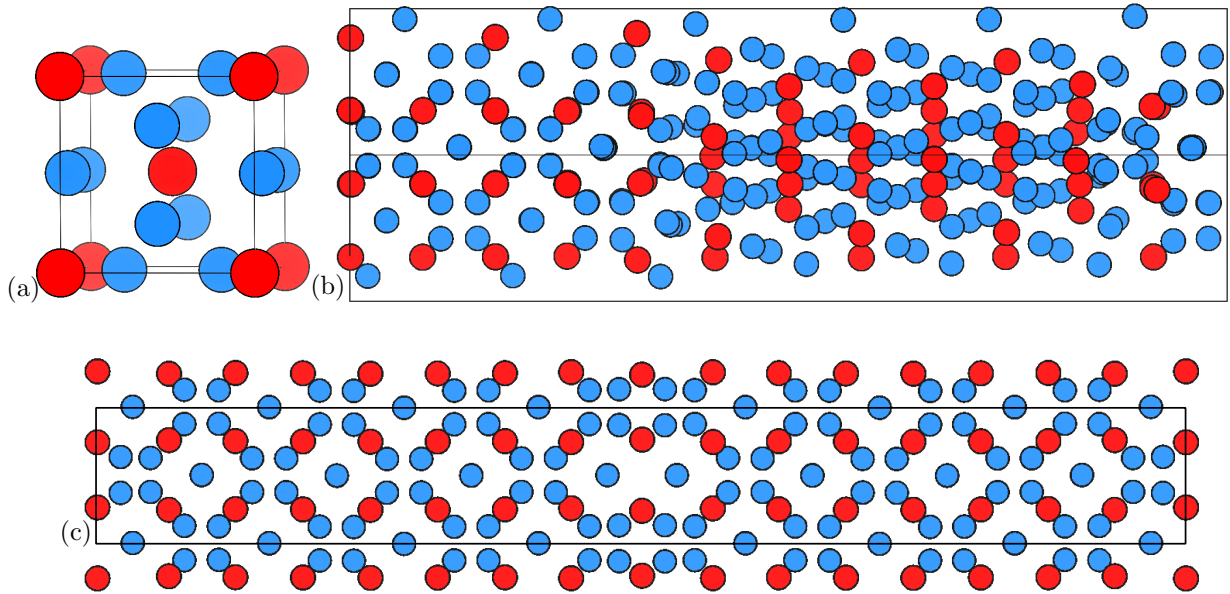
\begin{figure*}[t]
\centering
    \def\svgwidth{\textwidth}
\begingroup%
  \makeatletter%
  \providecommand\color[2][]{%
    \errmessage{(Inkscape) Color is used for the text in Inkscape, but the package 'color.sty' is not loaded}%
    \renewcommand\color[2][]{}%
  }%
  \providecommand\transparent[1]{%
    \errmessage{(Inkscape) Transparency is used (non-zero) for the text in Inkscape, but the package 'transparent.sty' is not loaded}%
    \renewcommand\transparent[1]{}%
  }%
  \providecommand\rotatebox[2]{#2}%
  \newcommand*\fsize{\dimexpr\f@size pt\relax}%
  \newcommand*\lineheight[1]{\fontsize{\fsize}{#1\fsize}\selectfont}%
  \ifx\svgwidth\undefined%
    \setlength{\unitlength}{583.27645638bp}%
    \ifx\svgscale\undefined%
      \relax%
    \else%
      \setlength{\unitlength}{\unitlength * \real{\svgscale}}%
    \fi%
  \else%
    \setlength{\unitlength}{\svgwidth}%
  \fi%
  \global\let\svgwidth\undefined%
  \global\let\svgscale\undefined%
  \makeatother%
  \begin{picture}(1,0.4802164)%
    \lineheight{1}%
    \setlength\tabcolsep{0pt}%
    \put(0,0){\includegraphics[width=\unitlength,page=1]{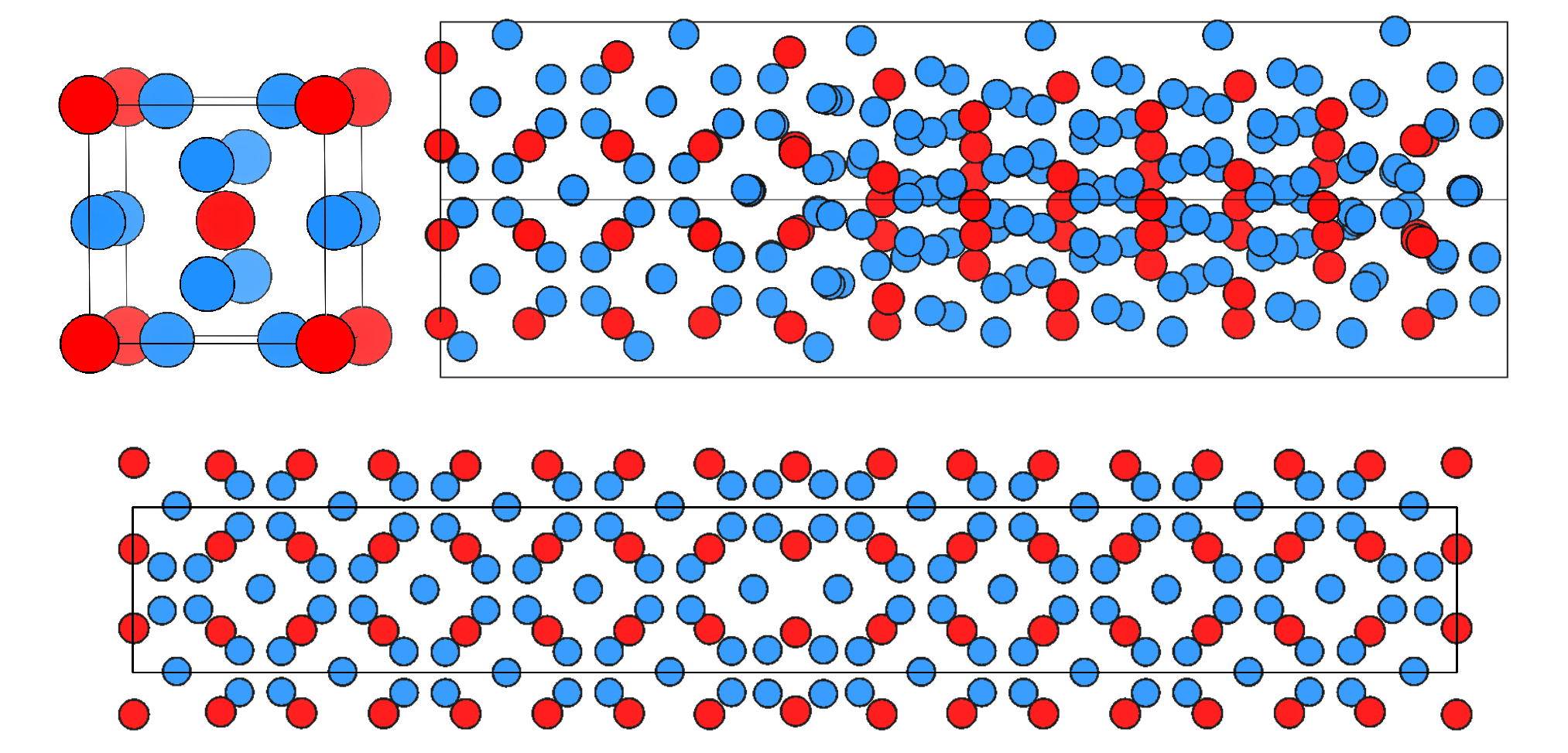}}%
    \put(0.05650948,0.05401518){\color[rgb]{0,0,0}\makebox(0,0)[lt]{\lineheight{1.25}\smash{\begin{tabular}[t]{l}(c)\end{tabular}}}}%
    \put(0.01520836,0.24327226){\color[rgb]{0,0,0}\makebox(0,0)[lt]{\lineheight{1.25}\smash{\begin{tabular}[t]{l}(a)\end{tabular}}}}%
    \put(0.25074957,0.24460142){\color[rgb]{0,0,0}\makebox(0,0)[lt]{\lineheight{1.25}\smash{\begin{tabular}[t]{l}(b)\end{tabular}}}}%
  \end{picture}%
\endgroup%

\caption{\label{fig:110atoms} Atomic structures with niobium colored in blue and tin in red. (a) The unit cell of cubic A15 Nb$_3$Sn with the tin atoms forming a bcc-lattice and chains of niobium atoms running along the faces of the cubes in the three orthogonal directions. Relaxed atomic positions in (b) the (110)-twist boundary cell and (c) the (110)-tilt boundary cell reported in table \ref{tab:GBs}.}
\end{figure*}

\section{Methods}
We perform density-functional theory (DFT) calculations within the pseudopotential framework using open-source plane wave software JDFTx~\cite{Arias1992, Sundararaman2017}. The electronic states are calculated for the outer electrons of niobium ($4p^6 5s^2 4d^3$) and tin ($4d^{10} 5s^2 5p^2$) while the atomic cores are treated using ultrasoft pseudopotentials~\cite{Garrity2014a}. We use the Perdew-Burke-Ernzerhof (PBE) approximation to the exchange-correlation functional and employ a 12 Hartree planewave cutoff energy~\cite{Perdew1996a}. Defect energies are calculated at high electronic temperatures using a Fermi function with a 5~milliHartree electronic temperature, chosen to be close to the experimental growth temperature of Nb$_3$Sn cavities~\cite{Lee2019, Posen2017, Porter2018}. Parameters relevant to our superconductivity analysis such as the densities of states are calculated at zero-temperature using the cold-smearing method developed by Marzari where we again use a smearing width of 5~mH to maintain the same tolerance with respect to k-point sampling~\cite{Marzari1999}. With these parameters, we calculate the lattice parameter for cubic A15 Nb$_3$Sn to be 5.271~\AA, in excellent agreement to its measured value of 5.289~\AA~\cite{Godeke2006, Devantay1981}. For cubic A15 Nb$_3$Sn we sample $6^3$ k-points in the Brillouin zone then transform to a maximally localized Wannier function basis to perform a dense Monte Carlo sampling to accurately calculate the density of states~\cite{Marzari1997}. The k-point meshes for the grain-boundary cells are chosen to have a sampling density comparable to the unit cell calculation, and their densities of states are calculated with a tetrahedral interpolation scheme. All results of the boundary cells involve fully relaxed internal atomic coordinates. We also consider the effects of lattice relaxation by setting the interfacial plane at the bulk lattice constant and allowing the lattice to relax along the boundary plane normal. We note, however, that the lattice relaxations do not significantly change any of the boundary energies so the results reported in this paper are calculated at the bulk lattice constant.

\section{Boundary structures and energies}
Nb$_3$Sn is an intermetallic alloy in the A15 phase; its cubic unit cell is displayed in figure \ref{fig:110atoms}(a)~\cite{Momma:db5098}. The A15 structure is characterized with one species of atom (tin) forming a bcc-lattice and each cube face containing two of the other atomic species (niobium). The atoms on each cube face form long one-dimensional chains of transition metals in the three orthogonal directions~\cite{King1990}. This particular structure is characteristically accompanied with a high Fermi-level density of states~\cite{Devantay1981}.

The electronic structure and properties of the A15 superconductors are known to be extremely sensitive to point defects, particularly when the structure strays away from its ideal stoichiometry~\cite{Sitaraman2019, Kate2017}. Here, we investigate the impact of extended defects on the properties of Nb$_3$Sn by performing DFT calculations on grain boundaries, including boundaries containing point defects. To our knowledge, the only existing first-principles studies on grain boundaries in this material are an unpublished thesis \cite{BYRNE2017} and a recent study that looked at three defect sites in one boundary cell but included no results on the electronic structure \cite{Lee2020}.

A grain boundary refers to the interface between two crystal grains of differing orientations or compositions. Geometrically, grain boundaries can be described by the interfacial plane, axis of rotation, and misorientation angle between the two grains~\cite{SMALLMAN2014415}. The two idealized boundary types are the twist and tilt boundaries, and an example of each is shown in figure~\ref{fig:110atoms}(b) and (c) respectively.

Grain boundaries are planar defects with excess free energy per unit area, corresponding to the energy to form the surface defect. For a simple estimate of a grain boundary energy, we follow Ref.~\citenum{Rohrer2011} to estimate the amount of elastic work done to create a surface based on the material's elastic modulus. Since Nb$_3$Sn has a elastic modulus of $\sim$127~GPa \cite{Hojo2006}, we expect grain boundary energies roughly in the range of 700--1500~mJ~m$^{-2}$.

\begin{table}
\caption{\label{tab:GBs}List of boundaries labeled with their interfacial plane and boundary type, misorientation angle, distance between boundaries, and relaxed grain boundary energy, $\gamma$.}
 \begin{indented}
 \item[]\begin{tabular}{@{}llll} 
 \br
Boundary  & $\theta$ & Grain Separation  & $\gamma$ (mJ~m$^{-2}$) \\
 \mr
  (110)-tilt & 90$^{\circ}$& 2.99 nm & 1455 \\ 

 (112)-tilt  & 70.5$^{\circ}$ & 2.59 nm & 840 \\

 (120)-tilt  & 53.1$^{\circ}$ & 2.36 nm & 1440\\

 (100)-twist  & 90$^{\circ}$ & 2.64 nm & 650 \\

 (110)-twist  & 70.5$^{\circ}$& 2.36 nm & 1300 \\
 \br
\end{tabular}
\end{indented}
\end{table}

Table \ref{tab:GBs} lists a selection of relaxed tilt and twist grain boundary cells with the fourth column indicating the surface formation energy of each boundary. The amount of energy to form a surface defect from a pure bulk crystal is determined by the energy difference between the total energy of a cell containing the surface defect and the total energy of a bulk cell containing the same number of atoms. The grain boundary energy, $\gamma$, is this energy difference divided by the area of the interfacial plane to become an intensive quantity
\begin{equation}
\gamma = \frac{(E_{\textrm{GB}} - E_{\textrm{bulk}})}{2A},
\end{equation}
where $E_{\textrm{GB}}$ and $E_{\textrm{bulk}}$ refer to the total energies of the boundary structure and the bulk structure of the same size, and $A$ is the area of one of the two interfaces contained within a surface defect unit cell. All table entries except for the (100)-twist involve symmetric structures, meaning the uniform bulk crystal is restored with a vanishing misorientation angle \cite{Runnels2016}. In the case of the (100)-twist, two niobium atoms overlap near the boundary plane and one niobium atom removed at each interface is thereby removed. A chemical potential enters the surface energy calculation to account for the change in stoichiometry, and we use the chemical potential of bulk niobium as Nb$_3$Sn is often grown on a niobium substrate.

The structures in table~\ref{tab:GBs} all fall within the expected energy range, and we have succeeded in identifying structures near the low end of the expected energy range for both the tilt and twist classes. These low energy boundaries provide representative examples of high-angle boundaries that we would expect to find in realistic samples. Understanding grain boundary structures and energies can advance applications of this material as grain boundary energies are one of the key factors driving grain growth, and larger grains produce weaker flux pinning force densities \cite{Livingston1977, Gowacki1988, Xu2017}.

\section{The impact of a clean boundary on electronic structure}
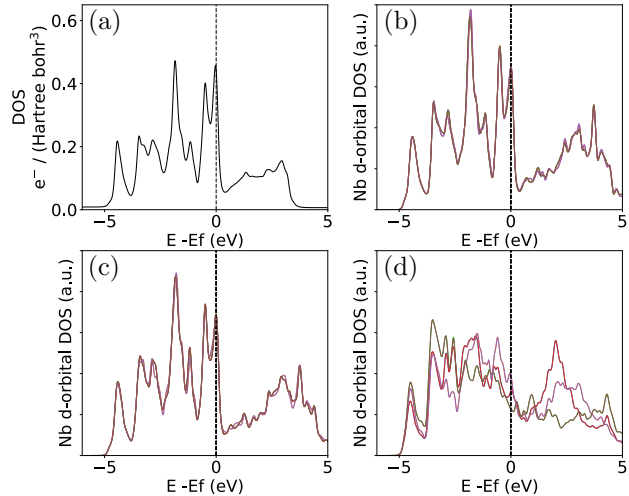
\begin{figure}[t]
    \def\svgwidth{\columnwidth}
\begingroup%
  \makeatletter%
  \providecommand\color[2][]{%
    \errmessage{(Inkscape) Color is used for the text in Inkscape, but the package 'color.sty' is not loaded}%
    \renewcommand\color[2][]{}%
  }%
  \providecommand\transparent[1]{%
    \errmessage{(Inkscape) Transparency is used (non-zero) for the text in Inkscape, but the package 'transparent.sty' is not loaded}%
    \renewcommand\transparent[1]{}%
  }%
  \providecommand\rotatebox[2]{#2}%
  \newcommand*\fsize{\dimexpr\f@size pt\relax}%
  \newcommand*\lineheight[1]{\fontsize{\fsize}{#1\fsize}\selectfont}%
  \ifx\svgwidth\undefined%
    \setlength{\unitlength}{904.8717652bp}%
    \ifx\svgscale\undefined%
      \relax%
    \else%
      \setlength{\unitlength}{\unitlength * \real{\svgscale}}%
    \fi%
  \else%
    \setlength{\unitlength}{\svgwidth}%
  \fi%
  \global\let\svgwidth\undefined%
  \global\let\svgscale\undefined%
  \makeatother%
  \begin{picture}(1,0.80030678)%
    \lineheight{1}%
    \setlength\tabcolsep{0pt}%
    \put(0,0){\includegraphics[width=\unitlength,page=1]{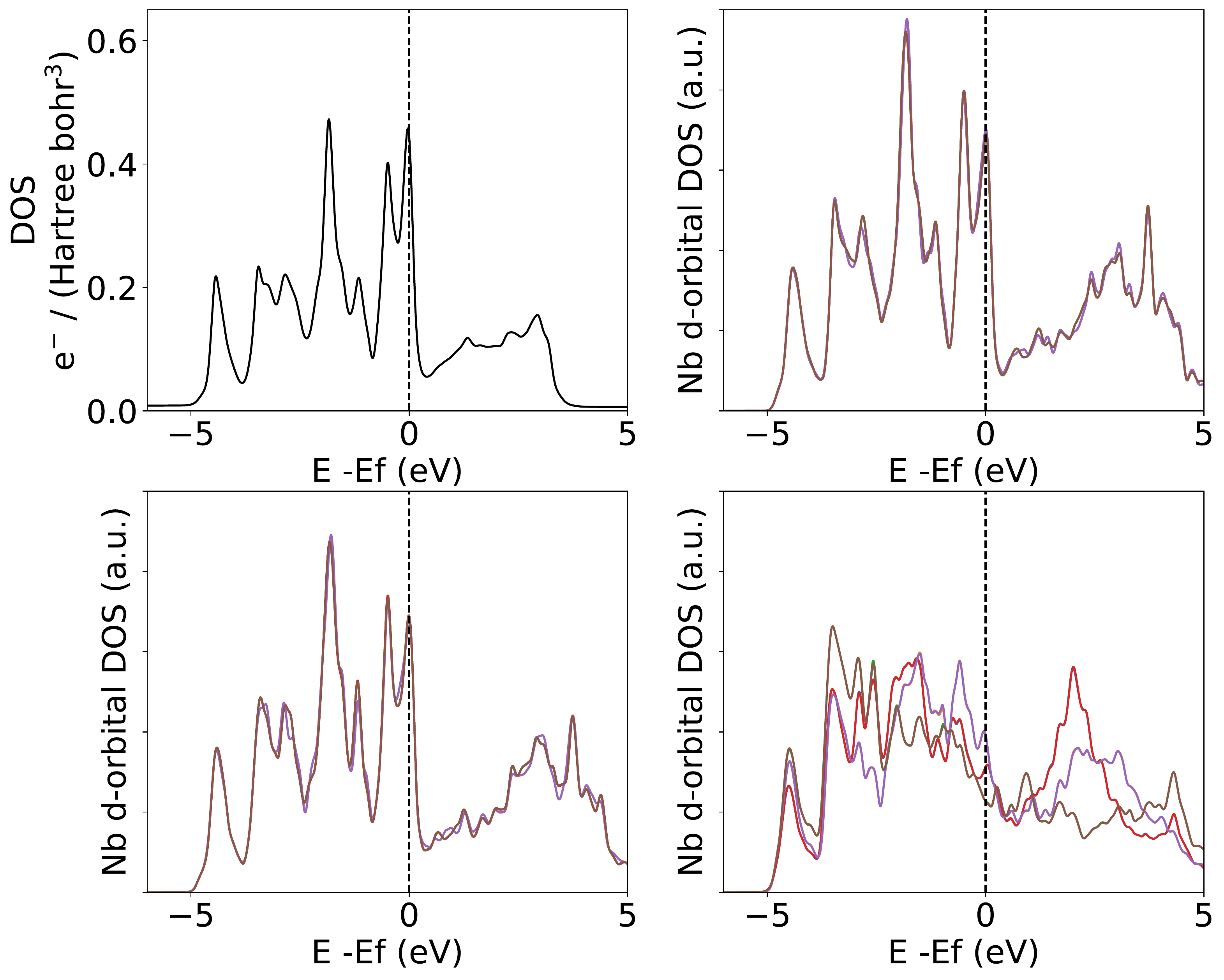}}%
    \put(0.12956887,0.75321615){\color[rgb]{0,0,0}\makebox(0,0)[lt]{\lineheight{1.25}\smash{\begin{tabular}[t]{l}(a)\end{tabular}}}}%
    \put(0.60169457,0.75387163){\color[rgb]{0,0,0}\makebox(0,0)[lt]{\lineheight{1.25}\smash{\begin{tabular}[t]{l}(b)\end{tabular}}}}%
    \put(0.12926725,0.36061651){\color[rgb]{0,0,0}\makebox(0,0)[lt]{\lineheight{1.25}\smash{\begin{tabular}[t]{l}(c)\end{tabular}}}}%
    \put(0.60113661,0.35996755){\color[rgb]{0,0,0}\makebox(0,0)[lt]{\lineheight{1.25}\smash{\begin{tabular}[t]{l}(d)\end{tabular}}}}%
  \end{picture}%
\endgroup%

\caption{\label{fig:dNb} (a) Total density of states in bulk Nb$_3$Sn calculated in a maximally localized Wannier function basis. The d-orbital projected density of states of niobium atoms plotted on the same scale in (b) a pure bulk cell, and in the (c) bulk region and (d) grain boundary core of (112)-tilt boundary cell. Each of (b--d) are plotted on the same scale for direct comparison.}
\end{figure}
A notable feature of the A15 superconductors is their high Fermi-level density of states. For weakly-coupled superconductors, the critical temperature can be described with the BCS equation, which exhibits an exponential dependence on the Fermi-level density of states $N(0)$
\begin{equation}
\label{eq:BCS}
    k_{\textrm B} T_{\textrm c} = 1.14 E_{\textrm D} \textrm e^{-1/N(0)V},
\end{equation}
and $E_\textrm D$ refers to the cutoff energy for phonons in the Debye model. Strongly-coupled superconductors such as Nb$_3$Sn are more accurately described with Eliashberg theory applied within a density-functional framework to calculate the electron-phonon coupling, $V$ from first principles \cite{eliashber, Brown2016a, Brown2016}. The critical temperature in this case is determined with the phenomenological McMillan formula, which  exhibits the same exponential dependence on $N(0)$ as equation~(\ref{eq:BCS}) \cite{McMillan1968}.

Figure \ref{fig:dNb}(a) shows the total density of states of cubic A15 Nb$_3$Sn. To visualize the contribution of the 1D niobium chains, we plot the partial projected density of states along d-orbitals of bulk niobium atoms in figure~\ref{fig:dNb}(b). 

The high peak in the density of states at the Fermi-level varies rapidly on small energy scales, dropping by more than a factor of 2 within 0.2~eV. This feature is a result of the conducting d-orbitals of the long 1D-chains of niobium atoms, or other transition metals for the other A15 superconductors~\cite{Godeke2006, Devantay1981}. This is confirmed by comparing figure~\ref{fig:dNb}(a) and (b), revealing the direct correspondence between the niobium atom d-orbital projected density of states with the total density of states.

Given the strong dependence superconductivity has on the Fermi-level density of states, especially for the A15 superconductors, we investigate how grain boundaries impact the density of states. In figure~\ref{fig:dNb} we plot the d-orbital projected density of states of (c) six niobium atoms in the bulk region and (d) six niobium atoms on the grain boundary core in the (112)-tilt boundary. We note that figures \ref{fig:dNb}(b--d) are plotted on the same scale and can be compared relative to one another. The niobium atoms in the bulk region of the boundary cell have densities of states nearly indistinguishable from that of niobium atoms in infinite bulk Nb$_3$Sn, showing that our cell is converged with respect to the boundary separation. For the niobium atoms on the grain-boundary core, the Fermi-level density of states is significantly reduced, roughly by a factor of two. A global reduction of this magnitude would decrease the superconducting transition temperature by over 10~K, a significant portion of its experimentally measured value of 18~K.

\begin{figure}[t]
\includegraphics[width=\columnwidth]{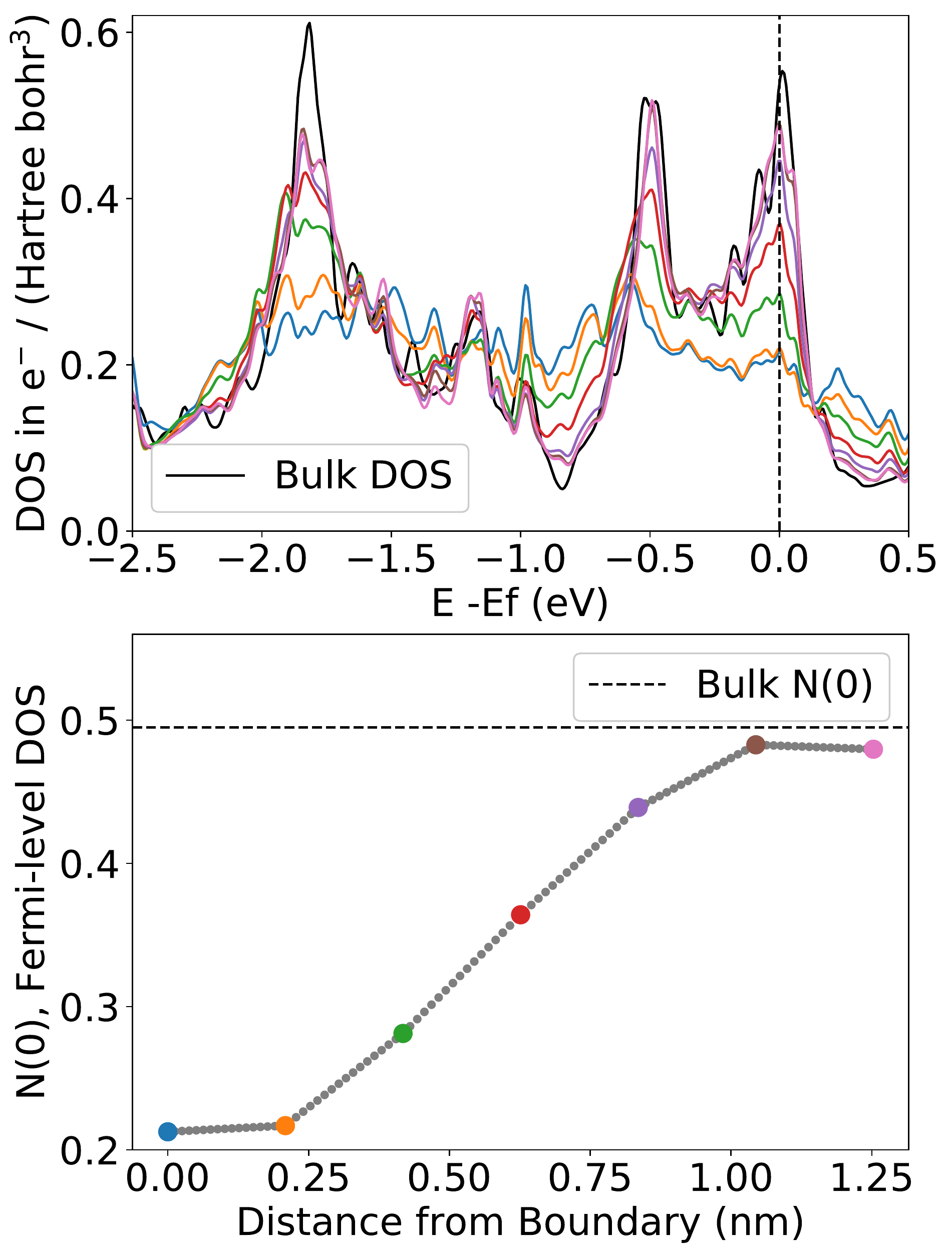}
\caption{\label{fig:localDOS} Local density of states curves across slabs in the (112)-tilt boundary cell (top). The corresponding Fermi-level density of states profile as a function of distance from the boundary (bottom).}
\end{figure}

To see how far this reduction extends into the crystal, we perform local density of states calculations, averaging over slabs perpendicular to the boundary plane. Results from this local density of states analysis and the corresponding profile of the Fermi-level density of states are plotted in figure~\ref{fig:localDOS} for the (112)-tilt boundary. We find that the factor of two reduction in the Fermi-level density of states at the grain boundary core decays smoothly back to the bulk value, in $\sim$1~nm from this boundary, and in $\sim$1--1.5~nm from boundaries in general. This range of impact of the boundary on the electronic structure is surprisingly wide, a fact which we will see has important implications for impurities and the impact of boundaries on $T_c$.

\section{Interactions between boundaries and point defects}
\subsection{Binding energy of tin antisite defects to a boundary}
Understanding how defects interact with grain boundaries in Nb$_3$Sn can advance growth techniques and superconducting applications of this material. The diffusivity of tin atoms through grain boundaries relative to bulk is important for Nb$_3$Sn growth~\cite{Osamura1986}. Tin defects at grain boundaries have been observed to be particularly detrimental to the material's superconducting properties across multiple applications~\cite{Xu2017, Sandim2013, Lee2020}. Accordingly, we first turn our attention towards tin defects near a grain boundary.
\begin{figure}[t]
\includegraphics[width=\columnwidth]{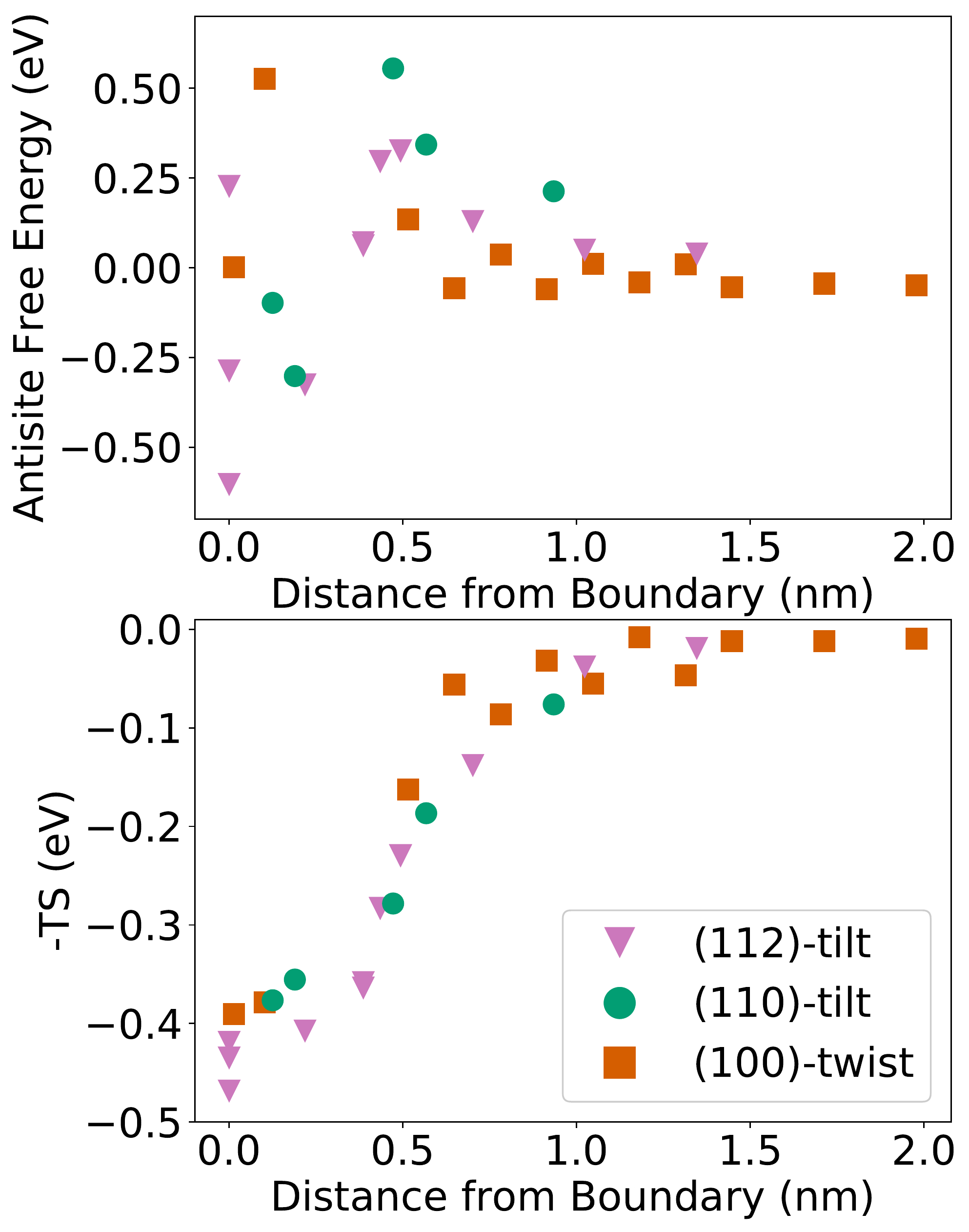}
\caption{\label{fig:antisite} Binding free-energies of Sn$_\textrm{Nb}$ defects to grain boundaries (top) and the corresponding electronic entropy contributions to the free energy (bottom) calculated at a high electronic temperature of 5~milliHartree comparable to experimental growth temperatures.}
\end{figure}

The prevailing tin defect in A15 Nb$_3$Sn consists of a tin atom sitting on niobium site, denoted as Sn$_\textrm{Nb}$~\cite{Besson2007}. In the top panel of figure~\ref{fig:antisite}, we plot the defect free-energies of tin substitutions relative to a tin substitution in bulk as a function of distance from a boundary in the (110)-tilt, (112)-tilt, and (100)-twist boundary cells. The bottom panel of figure~\ref{fig:antisite} displays the electronic entropy component of these free energies.

We find a strong, extended electronic interaction between tin antisite defects and grain boundaries that decays to bulk behavior within $\sim$1--1.5~nm. Near the boundary, we find a range of attractive and repulsive sites in each of the three boundary cells and conclude that the trends in these antisite free-energies may be complicated by structural details of the boundaries. Remarkably, however, the electronic entropy components of these free energies collapse onto a single curve when plotted as a function of distance. The resulting curve explains the general magnitude and range of the interaction with the boundary. We find significant free-energy gains approaching $\sim$0.5~eV favoring segregation of tin defects towards grain boundaries, providing an important gettering effect that could be exploited in the production of the superconducting material to control how tin distributes within various regions throughout the crystalline matrix.

\subsection{Impact of substitutional defects on a boundary's electronic structure}
To gain insight on the combined effect of defects and grain boundaries on superconductivity in this material, we look to see how various defects impact the local Fermi-level density of states. In light of the nearly universal electronic interactions calculated above, we consider for these investigations a single example for a case study, selecting the (112)-tilt boundary which contains an interfacial plane that has been experimentally observed~\cite{Lee2020}.

Because of the prevalence of metallic substitutions detected at grain boundaries in Nb$_3$Sn~\cite{Suenaga1983, Sandim2013, Lee2020}, we consider the two antisite defects Sn$_{\textrm{Nb}}$ and Nb$_{\textrm{Sn}}$ as well as defects of common ternary additions Ti$_{\textrm{Nb}}$, Ta$_{\textrm{Nb}}$, and Cu$_{\textrm{Sn}}$~\cite{Livingston1977, Tafto1984}. For a defect on the grain boundary core, we put the defect atom on the lowest-energy defect site identified within the core of the (112)-tilt boundary. We then compare the effect of this substitutional defect to the same defect when instead placed in the bulk region in a second calculation.

\begin{figure}[t]
    \def\svgwidth{\columnwidth}
    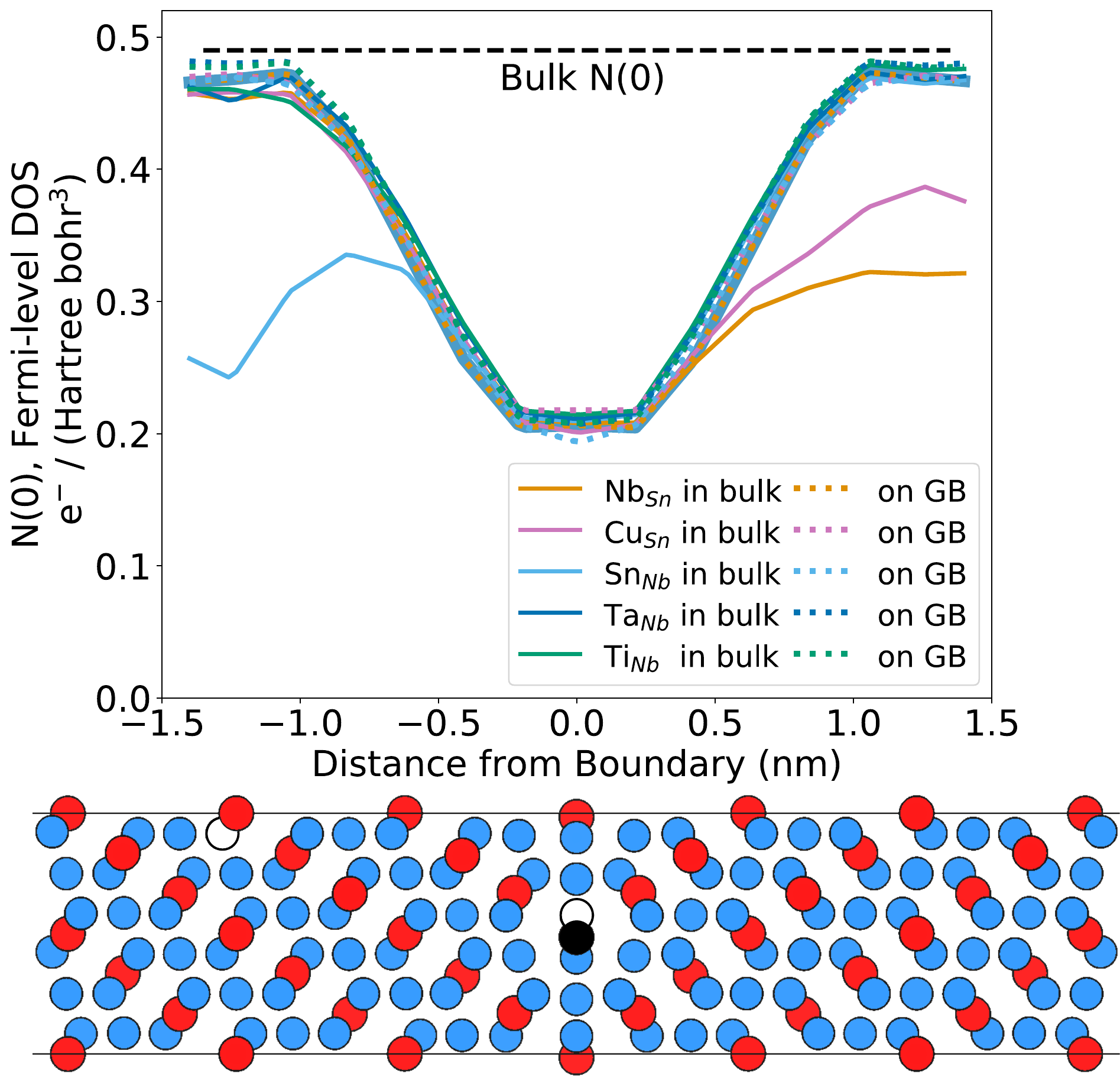
\caption{\label{fig:DOSimpact}(a) Local Fermi-level density of states across the (112)-tilt boundary including one defect atom on the grain boundary core (dotted lines) or in bulk (solid lines). (b) Atomic configuration displaying the tin and niobium atoms (red, blue), and the tin and niobium sites for substitutional defects (black, white).}
\end{figure}

Above, we investigated the effect of a clean grain boundary on the local Fermi-level density of states. Here, we repeat the same procedure in the presence of one substitutional defect. Figure~\ref{fig:DOSimpact}(a) compares the profiles of the Fermi-level density of states resulting from Sn$_{\textrm{Nb}}$, Nb$_{\textrm{Sn}}$, Ti$_{\textrm{Nb}}$, Ta$_{\textrm{Nb}}$, and Cu$_{\textrm{Sn}}$ defects, and figure~\ref{fig:DOSimpact}(b) indicates the substitution sites that we use in this study. The profile of the clean grain boundary free of point defects is displayed in figure~\ref{fig:localDOS}.

We find that, because a clean grain boundary itself already degrades the Fermi-level density of states by a factor of two, placing a point defect on the grain boundary core does not degrade it further. In contrast, placing defects in the bulk region allows for a greater impact on the local Fermi-level density of states. The most notable reductions arise from the Sn$_\textrm{Nb}$, Nb$_{\textrm{Sn}}$, and Cu$_{\textrm{Sn}}$ defects in bulk. These particular point defects profoundly disturb the conducting d-orbitals along the niobium chains. Interestingly, the Ti$_\textrm{Nb}$ and Ta$_\textrm{Nb}$ defects in bulk preserve most of the Fermi-level density of states, likely due to the chemical similarity of these elements to niobium, consistent with empirical observations that these elements do not degrade, and even can slightly enhance, superconducting performance~\cite{Suenaga1986}.

The disparity between the effects from these defects underscores the importance of distinguishing variations in stoichiometry arising from ternary elements from intrinsic variations in tin content which can have entirely different consequences for the superconductivity of the material. Furthermore, the strong dependence on the impurities' positions indicates the importance of high-resolution atomic imaging to measure atomic composition profiles and accurately determine the spatial extent of defect segregation.

\section{Superconductivity and grain boundary composition}
\begin{figure*}[t]
\centering
    \def\svgwidth{.96\textwidth}
\begingroup%
  \makeatletter%
  \providecommand\color[2][]{%
    \errmessage{(Inkscape) Color is used for the text in Inkscape, but the package 'color.sty' is not loaded}%
    \renewcommand\color[2][]{}%
  }%
  \providecommand\transparent[1]{%
    \errmessage{(Inkscape) Transparency is used (non-zero) for the text in Inkscape, but the package 'transparent.sty' is not loaded}%
    \renewcommand\transparent[1]{}%
  }%
  \providecommand\rotatebox[2]{#2}%
  \newcommand*\fsize{\dimexpr\f@size pt\relax}%
  \newcommand*\lineheight[1]{\fontsize{\fsize}{#1\fsize}\selectfont}%
  \ifx\svgwidth\undefined%
    \setlength{\unitlength}{340.38366765bp}%
    \ifx\svgscale\undefined%
      \relax%
    \else%
      \setlength{\unitlength}{\unitlength * \real{\svgscale}}%
    \fi%
  \else%
    \setlength{\unitlength}{\svgwidth}%
  \fi%
  \global\let\svgwidth\undefined%
  \global\let\svgscale\undefined%
  \makeatother%
  \begin{picture}(1,0.70099898)%
    \lineheight{1}%
    \setlength\tabcolsep{0pt}%
    \put(0,0){\includegraphics[width=\unitlength,page=1]{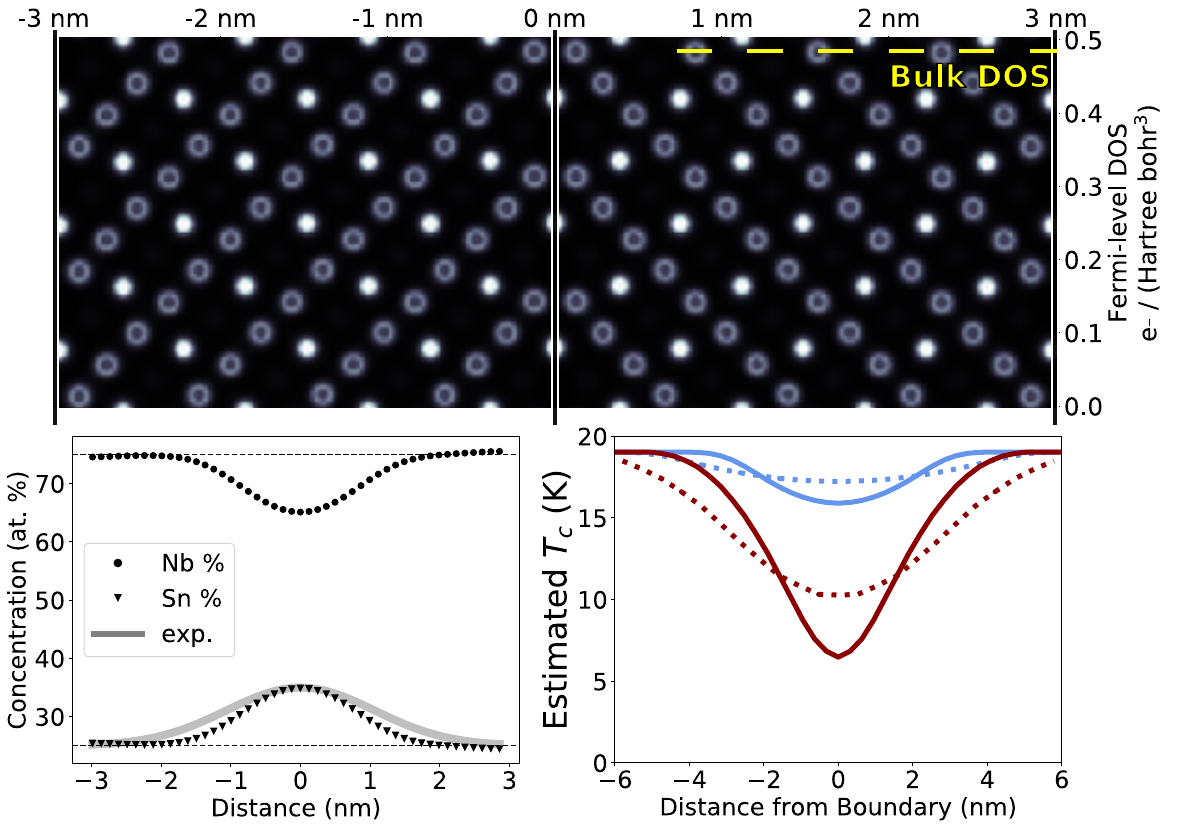}}%
    \put(0.00977937,0.35846041){\color[rgb]{0,0,0}\makebox(0,0)[lt]{\lineheight{1.25}\smash{\begin{tabular}[t]{l}(a)\end{tabular}}}}%
    \put(0.00950871,0.05367053){\color[rgb]{0,0,0}\makebox(0,0)[lt]{\lineheight{1.25}\smash{\begin{tabular}[t]{l}(b)\end{tabular}}}}%
    \put(0.94117873,0.35846041){\color[rgb]{0,0,0.00392157}\makebox(0,0)[lt]{\lineheight{1.25}\smash{\begin{tabular}[t]{l}(c)\end{tabular}}}}%
    \put(0.90224779,0.05367053){\color[rgb]{0,0,0}\makebox(0,0)[lt]{\lineheight{1.25}\smash{\begin{tabular}[t]{l}(d)\end{tabular}}}}%
    \put(0,0){\includegraphics[width=\unitlength,page=2]{figure6a_6d.pdf}}%
  \end{picture}%
\endgroup%

\caption{\label{fig:SnConc} (a) Two-dimensional slice of the electron density in the (110)-tilt boundary. (b) Atomic concentration profile added to the (110)-tilt boundary to simulate the interfacial excess measurements reported in Ref.~\citenum{Lee2020}, approximated by the gray curve. (c) Local Fermi-level density of states profiles around the clean and tin-rich (110)-tilt boundaries. (d) Resulting estimates of a local superconducting transition temperature $T_\textrm{c}$ obtained by averaging the profiles over the volume of a sphere. Solid lines represent an average over the expected size of a Cooper pair in this material and dotted lines display how this result changes with size.}
\end{figure*}

Finally, we can put the above results together to produce a model of how  grain boundary composition affects the local superconducting properties in Nb$_3$Sn. A popular model to describe flux pinning at grain boundaries uses diffuse isotropic scattering to estimate how the electronic mean free path is reduced near a boundary, likewise reducing the coherence length and causing local variations of the superconducting properties \cite{Zerweck1981}. Given the sensitivity that $T_\textrm{c}$ in Nb$_3$Sn has to variations in the Fermi-level density of states, we expect these variations to be fundamental in determining the local superconducting properties. Accordingly, we present an alternative model of local superconductivity that may be better suited for Nb$_3$Sn and the other A15 superconductors.

Here, we study the impact of a grain boundary with tin concentrations corresponding to measurements of Ref.~\citenum{Lee2020} that reported a Gibbsian interfacial excess of 10--20 tin atoms/nm$^2$. To model their observations, we include tin defects at locations corresponding to the experimentally observed concentration profile within the (110)-tilt boundary cell displayed in figure~\ref{fig:SnConc}(a). Ref.~\citenum{Lee2020} reports a segregation width of $\sim$3~nm containing a maximum tin concentration of 35\%, and the gray curve in figure~\ref{fig:SnConc}(b) approximates this with a Gaussian displaying a peak concentration of 35\% and a width of $4\sigma=3$~nm. To ensure we do not overestimate the effects of tin-segregated grain boundaries, we add a modest excess of 10 tin atoms/nm$^2$ spanning a slightly smaller width than that reported. To confirm that our choice of defect locations for the excess tin atoms reproduces the experimental profile, figure~\ref{fig:SnConc}(b) shows our final atomic arrangement convolved with a Gaussian corresponding to the lateral resolution of the probe used in the aforementioned experiments. We have confirmed that our results are insensitive to rearrangement of the defects, provided the concentration profile remains the same.

The depression in the Fermi-level density of states from the tin-rich boundary extends much further than that of its stoichiometric counterpart, while the depth of the depression is only slightly deeper. To understand the impact of the clean and tin-rich boundary structures on the local superconducting transition temperature $T_\textrm{c}$, we note that first principles studies in Nb$_3$Sn have established that the primary impact on $T_\textrm{c}$ in this material comes from the Fermi-level density of states as opposed to other factors. Specifically, a first principles study on the impact on strain in Nb$_3$Sn showed that $\sim$80\% of the degradation of the superconducting properties from strain comes from the reduction in the Fermi-level density of states~\cite{Godeke2018}, and another first principles study on antisite defects in Nb$_3$Sn also found $T_\textrm{c}$ to be strongly correlated with the Fermi-level density of states~\cite{Sitaraman2019}. To provide local values for $T_\textrm{c}$, the Fermi-level density of states must be averaged over an appropriate length scale corresponding to the superconducting coherence length. Figure~\ref{fig:SnConc}(d) shows estimates for a local $T_\textrm{c}$ obtained by integrating the density of states profiles in figure~\ref{fig:SnConc}(c) over a sphere of radius 3~nm to coincide with the coherence length of Nb$_3$Sn. To explore the sensitivity of this result to the form of the averaging, figure~\ref{fig:SnConc}(d) also shows the result from averaging over a sphere of radius 5~nm.

We find, regardless of the form of the averaging, the reduction in $T_\textrm{c}$ to be wider and much deeper for the tin-rich boundary than for the clean boundary, although the magnitude of the effect is sensitive to averaging. The local $T_\textrm{c}$ around a clean grain boundary is barely degraded because the depression in the density of states has a diameter of $\sim$2~nm while a Cooper pair has a radius of 3~nm. However, a boundary filled with tin-defects widens the depression in the density of states, to the detriment of the local superconducting properties: the reduction in $T_\textrm{c}$ becomes significantly wider and deeper. This result correlates well with the findings of Ref.~\citenum{Lee2020}, reporting high quality factors from Nb$_3$Sn SRF cavities containing clean grain boundaries and degraded quality factors from cavities containing tin-rich grain boundaries. We predict similar degradations in the local superconducting properties from boundaries filled with Nb$_{\textrm{Sn}}$ and Cu$_{\textrm{Sn}}$ defects, and note that similar effects on copper's influence have been observed \cite{Rodrigues2007}. 

Grain boundaries with low $T_\textrm{c}$ are a candidate mechanism that lowers the first vortex entry field in Nb$_3$Sn SRF cavities, and enhancing flux pinning in Nb$_3$Sn superconducting wires is the only opportunity to improve critical current densities \cite{Pack2020, Xu2017}. Ginzburg-Landau simulations can build off of these local $T_\textrm{c}$ estimates to further our understanding of flux penetration and pinning at grain boundaries in Nb$_3$Sn and how these processes are affected by grain boundary composition~\cite{Transtrum2011, Li2017, Carlson2020}.

\section{Conclusions}
This paper presents the first \emph{ab initio} investigation of the physics of different boundary types in Nb$_3$Sn. To our knowledge, this study is the first to consider twist boundaries in addition to tilts and to include boundary planes with distinct orientations. We present a selection of energetically favorable structures and identify structures at the low end of the expected energy range for both boundary types. These representative structures provide a foundation to examine the physics of grain boundaries in superconducting Nb$_3$Sn broadly.

We present the impact of multiple classes of grain boundaries on the electronic structure and find that the Fermi-level density of states is significantly degraded, which is expected from the disrupted niobium chains in the A15 phase. A global reduction of this amount would decrease $T_\textrm{c}$ by over 10~K. We find that grain boundaries have a striking long-range effect on the electronic structure, where the depression in the Fermi-level density of states extends out to $\sim$1--1.5~nm on each side. The narrowest depressions occur in the symmetric boundaries we study, widening slightly in the asymmetric cell we consider, and so we anticipate random high-angle grain boundaries in general to have depressions on the wider side of this range because of the greater disruption within the structure.

Looking towards the impact of a boundary's electronic structure on the binding energy of defects, we find a strong electronic interaction that also extends out {$\sim$1--1.5~nm}. Defect free-energies near the boundary are complicated by local structural details, but the electronic entropy contributions to the defect-boundary interaction remarkably collapse onto a single curve. This attractive electronic entropy contribution provides a ready explanation for the magnitude and range of the interaction favoring tin segregation at grain boundaries. We predict grain boundaries to have a full-width interaction range of $\sim${2--3~nm} with point defects, correlating well with atomic content measurements~\cite{Suenaga1983, Sandim2013, Lee2020}.

Defects also affect the electronic structure of grain boundaries. We find that point defects on the grain boundary core do not degrade the Fermi-level density of states much beyond the impact of the boundary itself. However, point defects away from the boundary core have much more of a relative impact, with bulk Sn$_\textrm{Nb}$, Nb$_\textrm{Sn}$, Cu$_\textrm{Sn}$ defects in particular being severely detrimental. In contrast, we find evidence that other defects such as Ti$_\textrm{Nb}$ and Ta$_\textrm{Nb}$ can be used as dopants at low concentration without severely impacting the electronic structure of the material. Electronic structure studies such as these can inform further exploration of prospective dopants without the cost and challenges of experimentally examining every possible dopant.

Finally, we introduce a novel model for a local superconducting transition temperature $T_\textrm{c}$ in the vicinity of a grain boundary that we expect to be applicable for Nb$_3$Sn and the other A15 superconductors. Given the fact that the transition temperature in Nb$_3$Sn is most sensitive to the Fermi-level density of states and the damaging impact Sn$_\textrm{Nb}$ defects have on this quantity, we contrast the effect of a clean grain boundary from the effect of a boundary containing a tin concentration profile analogous to those observed in experiment. We find a wide depression in the Fermi-level density of states from the tin-rich boundary that extends further than from a clean boundary of the same structure. We predict that this wider depression causes a large reduction in $T_\textrm{c}$ around the tin-rich boundary compared to the slight reduction in $T_\textrm{c}$ around a clean boundary. This result explains from first principles why grain boundaries with excess tin are significantly more detrimental to superconductivity than are clean grain boundaries and provides a fundamental understanding of how the control of Sn$_\textrm{Nb}$ defects becomes crucial for superconducting applications of Nb$_3$Sn.

Long standing efforts towards improving the performance of Nb$_3$Sn are resulting in a resurgence of the material, allowing for large scale superconducting applications to operate at higher fields and temperatures. Improving critical current densities in Nb$_3$Sn requires recognizing the relevant microstructures worth controlling. Producing finer grains without deleterious tin gradients is an ongoing challenge, and density-functional theory can potentially identify candidate additives to aid in experimental developments. In order to continue improving applications of Nb$_3$Sn, it is imperative to continue improving our microscopic description of competing defects within the material, and \emph{ab initio} grain boundary calculations provide a promising new perspective previously overlooked.

\ack 
We would like to thank Jim Sethna, David Muller, Matthias Liepe, and Ryan Porter of Cornell University and Sam Posen of Fermilab for insightful discussions regarding grain boundaries and Nb$_3$Sn, Jaeyel Lee and David Seidman of Northwestern University for sharing their experimental measurements, and Ravishankar Sundararaman of Rensselaer Polytechnic Institute for all the momentous first-principles software he continues to develop.

This work was supported by the US National Science Foundation under award PHY-1549132, the Center for Bright Beams.

\section*{References}

\bibliographystyle{myunsrt}
\bibliography{GrainBoundaryPaper}

\end{document}